\documentclass[twocolumn]{emulateapj}
\usepackage{amsmath}
\usepackage{rotating}
\usepackage{hyperref}
\usepackage{epsfig}

\begin{document}
\def\gapprox{\mathrel{\vcenter{\offinterlineskip \hbox{$>$}
    \kern 0.3ex \hbox{$\sim$}}}}
\def\lapprox{\mathrel{\vcenter{\offinterlineskip \hbox{$<$}
    \kern 0.3ex \hbox{$\sim$}}}}
\newcommand{\beq}{\begin{equation}}
\newcommand{\eeq}{\end{equation}}

\def\eps{\epsilon}
\def\epsem{\eps_{\rm EM}}
\def\epszero{\eps_{\rm 0}}
\def\epsrad{\eps_{\rm rad}}
\def\epsw{\eps_{\rm w}}
\def\epsj{\eps_{\rm jet}}
\def\epsii{\eps_{\rm II}}

\def\lnu{L_{\nu}}
\def\lbh{L_{\rm BH}}
\def\ldw{L_{\rm dw}}
\def\ledd{L_{\rm Edd}}
\def\Edotw{\dot{E}_{\rm w}}

\def\Msun{M_{\odot}}
\def\Mbh{M_{\rm BH}}
\def\Medd{M_{\rm Edd}}
\def\Macc{M_{\rm acc}}
\def\Minf{M_{\rm inf}}
\def\Moutf{M_{\rm outf}}

\def\mdot{\dot{m}}
\def\Mdotacc{\dot{M}_{\rm acc}}
\def\Mdotinf{\dot{M}_{\rm inf}}
\def\Mdotoutf{\dot{M}_{\rm outf}}
\def\Mdotbh{\dot{M}_{\rm BH}}
\def\Mdotedd{\dot{M}_{\rm Edd}}

\def\vw{v_{\rm w}}
\def\vwten{v_{\rm w,10}}

\def\pdot{\dot{p}_{\rm w}}

\def\Re{R_{\rm e}}
\def\rbh{R_{\rm BH}}

\shortauthors{Ostriker et al.}
\shorttitle{Momentum Driving of AGN feedback}

\title{Momentum Driving: which physical processes dominate AGN feedback?}

\author{Jeremiah P. Ostriker\altaffilmark{1,2}, Ena Choi\altaffilmark{1},
Luca Ciotti\altaffilmark{3}, Gregory S. Novak\altaffilmark{1}, and 
Daniel Proga\altaffilmark{4}}

\affil{$^1$Department of Astrophysical Sciences, Princeton University, 
        Princeton, NJ 08544, USA}
\affil{$^2$IoA, Cambridge, UK}
\affil{$^3$Department of Astronomy, University of Bologna, via Ranzani 1, 
        I-40127, Bologna, Italy}
\affil{$^4$Department of Physics and Astronomy, University of Nevada, 
        Las Vegas, NV, USA}

\begin{abstract}
The deposition of mechanical feedback from a supermassive black hole 
(SMBH) in an active galactic nucleus (AGN) into the surrounding galaxy
occurs via broad-line winds which must carry mass and radial momentum
as well as energy. The effect can be summarized by the dimensionless
parameter $\eta={\Mdotoutf}/{\Mdotacc}= {2 \epsw c^2}/{\vw^2}$ where
$\epsw$ ($\equiv \Edotw/(\Mdotacc c^2)$) is the efficiency by which 
accreted matter is turned into wind energy in the disc surrounding the 
central SMBH. The outflowing mass and momentum are proportional to 
$\eta$, and many prior treatments have essentially assumed that 
$\eta=0$. We perform one- and two-dimensional simulations and find 
that the growth of the central SMBH is very sensitive to the inclusion of 
the mass and momentum driving but is insensitive to the assumed 
mechanical efficiency. For example in representative calculations, the 
omission of momentum and mass feedback leads to an hundred fold 
increase in the mass of the SMBH to over $10^{10} \Msun$. When 
allowance is made for momentum driving, the final SMBH mass is much 
lower and the wind efficiencies which lead to the most observationally
acceptable results are relatively low with $\epsw \lesssim 10^{-4}$.

\end{abstract}

\section{Introduction}\label{intro}

Feedback from active galactic nuclei (AGNs) at the centers of galaxies 
is believed to have a significant effect on the evolution of those galaxies.  
However, the precise physical mechanisms by which this feedback 
occurs are greatly uncertain---perhaps more so than is commonly 
acknowledged. While much path-breaking and insightful work has been
done, it is also true that some of the most basic requirements, such as the 
necessity that mass, energy, and momentum be conserved, have not 
been imposed in several of the popular treatments of this subject. And the 
inclusion of the presently known and observed feedback processes is 
often treated selectively. The purpose of this paper is to attempt to lay 
out the physical framework for discussing the issues and to provide 
illustrative examples of the results obtained primarily from 
one-dimensional computations that include or exclude specific 
processes. We also include a treatment of the two-dimensional, 
axisymmetric case, presented in less detail, to show how the qualitative 
features carry over to this more realistic case. Definitive solutions are 
beyond present art in this field, so the focus will be on the qualitative 
features of the physical solutions rather than the detailed comparison 
with observations.

In outline, there are three phases to the overall phenomenon: 1) the 
provision of fuel for the central supermassive black hole (hereafter 
SMBH); 2) the generation of the outflowing stream of energy, mass and 
momentum from the vicinity of the SMBH; and 3) the absorption and 
transmission of this energy, mass and momentum by the ambient gas 
in the galaxy and the subsequent reactions of the ambient gas to that 
input.

1. The fueling is generally believed to be via infalling gas, and typically, 
two origins for that gas have been proposed; at high redshift ambient 
gas in discs liberated by the non-axisymmetric forces released during 
mergers is certainly important \citep{bar91}, while at lower redshift 
mergers fail and probably do not fuel AGNs \citep{li08}, but the 
processed gas released via normal stellar evolution provides an ample 
source \citep[][hereafter CO07]{mat83,shu83,cio91,pad93,cio07}. 
A primary clue as to which of these sources dominates in a specific case 
is provided by details of the metallicity distribution, since the 
reprocessed gas probably has super-solar metal abundance and 
will also show signs of stellar evolution such as higher nitrogen or 
S-process abundances. The clue to fueling by infalling globular clusters 
might be relatively low abundances of elements made in SN I such as Fe.
In almost all treatments a central disc mediates between the inflowing
material and the SMBH. Other sources, such as small stellar systems 
dragged in by dynamical friction have been considered from time to time. 
These stars, or others added to the central regions via loss-cone 
processes  \citep{beg80} can be shredded during tidal interactions with 
the central SMBH or by collisions with one another or with a central 
disc---the debris collecting in the disc and feeding the central SMBH via 
conventional mechanisms \citep{ost83,dai10}.

2. The outflows fall into three categories. The signature of AGNs is, of 
course, the enormous electromagnetic, luminous output, with major 
contributions from the IR bands to the gamma ray region. The bulk of 
the flux is typically in the ``UV bump" and the flux from this region thus 
dominates for the ``momentum driven winds" \citep[e.g., see][]{pro00,
kin03,pro04,deb09}. The region where this driving occurs is fairly close 
to the quasar  (50$\rbh \lesssim r \lesssim 500 \rbh$, where $\rbh$ is 
the SMBH Schwarzschild radius). However, the moderately hard X-rays 
determine the average photon energy: 
$\left< h \nu \right>={h \int \lnu \nu {\rm d}\nu}/{\int \lnu {\rm d}\nu}$ when 
integrated over the spectrum. This region of the spectrum dominates the 
photon heating, causing the heated gas to approach the mean photon energy: 
$1.5 k T_X = \left< h \nu \right>$ with $T_X \sim 2 \times 10^7$K \citep{saz04}. 
The resultant heating occurs over an extended range of radii: 100 pc$ 
<r< $3 kpc \citep[][hereafter Papers I and III]{cio09,cio10}, and it can be 
significant for $r \gtrsim$ 0.1 pc \citep{pro07}. It can efficiently drive 
outflows as shown in a series of papers by Ciotti, Ostriker and 
collaborators (cf.CO07 and references therein). For electromagnetic 
output there is, of course, no rest-mass component. The total energy 
emitted in this form has been established fairly accurately via the 
\citet{sol82} argument to be $\Delta E = \epsrad \Delta \Macc c^2$ 
with $\epsrad \sim 0.1-0.15$ \citep{yu02}. The momentum output, 
of course, is $\Delta p = \Delta E / c$.  In optically thick cases 
($\tau \gg 1$), the total momentum absorbed by the fluid can approach 
$\Delta p = \tau \Delta E / c$ \citep{deb09}. \citet{sil10} also consider
the importance of radiative momentum driven winds on galactic and 
cluster scales but limit the input to $L/c$, which can be considerably 
less than allowed in the optically thick case by \citet{cio07} or
\citet{deb09}.

Next, let us turn to mechanical output. Both broad- and narrow-line 
regions inject mass, energy and momentum into the surrounding gas, 
with the broad-line winds probably dominant. Since these are material 
flows with velocity in the vicinity of the SMBH, $\vw$, the mass outflow 
can be considerable. If we let the inflowing and outflowing mass rates 
be ($\Mdotinf$, $\Mdotoutf$), then conservation of mass, energy and 
momentum can be summed up with the following simple equations:
\beq
\Mdotacc = \Mdotinf - \Mdotoutf, \label{eq:Mdot}
\eeq
where $\Mdotacc$ is the mass rate actually accreted by the SMBH, and
\begin{subequations}
    \begin{eqnarray}
\Edotw &=&  \frac{1}{2} \Mdotoutf \vw^2\label{eq:edotw2}\\
      &=& \epsw \Mdotacc c^2, \label{eq:edotw1} 
    \end{eqnarray}
\end{subequations}
\beq
\pdot = \Mdotoutf \vw, \label{eq:pdot}
\eeq
are the wind energy and momentum, respectively. We have 
oversimplified matters by allowing only one wind velocity, when in fact 
Equation~\ref{eq:edotw2} requires $\left<\vw^2\right>$ and 
Equation~\ref{eq:pdot} requires $\left<\vw\right>$. Also, it is important
to specify exactly where and when the quantities in equations
\ref{eq:Mdot}--\ref{eq:pdot} are to be measured. In the conventional
treatment of this subject, the SMBH is surrounded by a disc or torus
to which matter has fallen from larger radius. Then, placing a sphere
around this disc or torus (at $r \sim 1$ pc), the instantaneous spherically
averaged infall through the sphere is $\Mdotinf(t)$ and the spherically
averaged outflow is $\Mdotoutf(t)$. The difference will be accreted onto
the SMBH unless driven out in disc originating winds; the latter of course
contributes to $\Mdotoutf$ and so the remainder $\Mdotinf - \Mdotoutf$
will be accreted. Two further complications are allowed for in some
detailed treatments: (i) the actual instantaneous value of $\Mdotacc$
is a time-lagged convolution of the quantity in Equation \ref{eq:Mdot}
since a finite time elapses as material is transported through the disc
to the central SMBH and (ii) star formation may (in fact frequently will)
occur in the disc, removing mass that would otherwise have accreted
onto the SMBH. Both of these complications are allowed for in CO07
and other work, neither is of dominant importance.

Now, defining the dimensionless ratio from Equation~\ref{eq:edotw2} 
and \ref{eq:edotw1} to be,
\beq
\eta \equiv \frac{\Mdotoutf}{\Mdotacc}= \frac{2 \epsw c^2}{\vw^2},
\label{eq:eta}
\eeq
we can now rewrite Equations~\ref{eq:Mdot}--\ref{eq:pdot} as
\begin{subequations}
    \begin{eqnarray}
\Mdotacc &=& \Mdotinf \frac{1}{1+\eta}\label{eq:Mdot_sol} ,\\
\Mdotoutf &=& \Mdotinf\frac{\eta}{1+\eta}, \\
\Edotw     &=& \frac{1}{2} \Mdotinf \vw^2 \frac{\eta}{1+\eta}
                         =\epsw \Mdotinf c^2 \frac{1}{1+\eta}, \\
\pdot        &=& \Mdotinf \vw \frac{\eta}{1+\eta}.
    \end{eqnarray} \label{eq:sol}
\end{subequations}
These Equations, \ref{eq:sol}a\--d, are, in fact, the ones that most
authors have adopted who treat AGN feedback as a unified process
comprising both infall and outflow. However, they typically 
adopt $\eta=0$, implicitly assuming $\vw \to \infty$, so that $\Mdotoutf$ 
and $\pdot$ are neglected and the two terms that are included, $\Edotw$ 
and $\Mdotacc$, may be overestimated. If it eventuated that $\eta$ really 
is a very small number, then not much error would be induced and one 
would be justified in neglecting the out-flowing mass and momentum 
and in setting $\Edotw \sim \epsw \Mdotinf c^2 $, as most authors 
assume. If we adopt for the efficiency of generating mechanical energy 
the value $\epsw = 5 \times 10^{-3}$, as done by \citet{spr05,joh09}, 
hereafter SDMH05 and JNB09 respectively, \citet{mcc10} and other authors, and we 
take $\vw = 10^4$ km/s ($\vwten$) \citep{moe09}, then we have from 
Equation~\ref{eq:eta}, $\eta = 9 \vwten^{-2}$ and all of the neglected 
effects may in fact be dominant; the bulk of the inflowing mass 
may be ejected in a broad-line disc wind, and the mass and momentum 
input deposited in the ambient gas may dominate over the energy input, 
which may be largely radiated away. Papers I and Paper III do include 
these effects, but do not spell out their significance. The principal 
purpose of the present paper is to do just that---to show, in specially 
simple one- and two-dimensional calculations the effects of including
or excluding mass, energy and momentum conservation when $\eta > 0$.
In addition to the papers referred to above which attempt to compute
both the infall to the central SMBH and the outflow from it in a unified
fashion, there are many others which postulate a central source and
then, after estimating the mass, momentum and energy flowing at of
that source (and some angular and temporal distribution thereof) do
effectively compute the effects of that injection of energy, mass and 
momentum onto the surrounding fluid. Space does not permit a
comprehensive description of this related subject of research, but
important papers include the following: cf. 
\citet{met94,ste08,fab09,ree09,ari10,gas10}.

The wind efficiency, $\epsw$, is not known very well---neither from 
observations nor from detailed physical simulations. But the best 
estimates from either of these sources might be in the range 
$1\times10^{-3} > \epsw > 3\times10^{-4}$ \citep{pro00,pro04,
kro07,sto09,kur09}, a factor of 5 to 17 smaller than the commonly 
adopted values and in a range where $\eta \lesssim 1$ if $\vwten 
\approx 1$.  A specific example may be useful.  \citet{moe09} study the 
quasar SDSS J0838+2955. They find a mechanical energy output of 
$4.5 \times 10^{45}$ erg/s, a mass outflow rate 10 times the accretion 
rate and a mechanical efficiency of $1 \times 10^{-3}$, and they quote 
other observational studies which indicate similar numbers.  From 
analyses of the ionization parameters in the broad-line winds,
estimates of the radial extent of the winds can be made; the above 
paper, and those quoted within indicate radii measured in 
kpc---consistent with the one-dimensional numerical work in Paper III. 

As shown in Papers I and III, an additional important question asks 
``what fraction of the sky is covered with the broad-line winds?" Again 
two approaches are possible. Empirically, on the order of 20-25\% of 
bright quasars show broad-line winds; this translates to $\sim \pi$ 
steradians or $\pi/2$ steradians in each conical outflow, if we assume 
that the wind is emitted symmetrically above and below the inner AGN 
disc. On the theory side, the radiation driven winds found by \citet{pro04},
via detailed hydro radiation-transfer calculations, cover $\sim \pi$ 
steradians, roughly consistent with the observational estimates.

Finally, let us turn to the narrow jets, the outflow observed from AGN in 
``radio mode", when the electromagnetic luminosity is considerably 
below the Eddington limit. M87 is an excellent nearby example of such 
a system. These are standard FRI radio sources. Here the jets are quite narrow
and appear to be comprised primarily of a relativistic fluid. The same type 
of calculation as presented in the last section would indicate that the 
out-flowing mass is of negligible importance and the energy output greatly 
dominates over the momentum output. The total energy output from 
these phases is considerable, but the accretion rates are thought to be 
low in these phases so the efficiencies of energy generation may be 
very high \citep[cf. for a computation][]{gam04}. Since so little 
mass is accreted in radio mode, the Soltan argument cannot be used 
to empirically estimate efficiencies, but, from the observational estimates 
of the energies available in the giant radio lobes, it may be that 
the AGN emits in radio mode considerably more energy than it does in 
wind mode. However the deposition from the intense but extremely 
narrow streams appears to be inefficient, and the jet drills through the 
gas in the surrounding galaxy, dumping most the energy into the 
intergalactic medium.  Thus, while it may act as the dominant feedback 
mechanism for the IGM (and we will return to this in a subsequent paper), 
it is of lesser importance than the radiative or wind components in 
heating and driving out the ambient gas from within a galaxy.

3. The interactions between the out-flowing energy, mass and momentum 
with the ambient fluids are complex and are just beginning to be studied 
with the needed detail. We focus here on the relatively gas poor elliptical
systems, since it is in these that the bulk of the mass in SMBHs is found,
The radiative interactions are perhaps easiest to 
describe. Since the mechanical momentum is conserved and cannot be 
radiated away, it can be a dominant effect. The minimum level of 
interaction is provided by electron scattering and, since the most 
luminous quasars are found to be clustered near the Eddington 
luminosity limit (at which level the momentum absorbed by electron 
scattering balances the gravitational force on the fluid from the central 
SMBH), we know that this effect is significant in many cases. 
Absorption of the out-flowing radiation will not, in general, reduce this 
effect, since typically the radiation is simply re-emitted in another band 
and electron scattering opacity is wavelength independent until the 
Klein-Nishina limit is reached at very high energies. In fact, in the optically 
thick limit, the radiation is transformed by dust absorption into the 
infra-red, but the effects in this case are even greater than in the simple 
case, since the scattering opacity of the dust to infra-red is, per atom, 
larger (by roughly a factor of 5) than the electron scattering cross section. 
For the bright ULIRGs, which may contain both an active AGN and a 
brighter starburst, there will be a near balance between the inward 
gravitational forces and the outward radiative momentum transfer 
on the dust \citep[cf.][CO07]{tho05}. Under these circumstances, the inner  
several hundred parsecs of the galaxy are analogous in their equilibrium 
structure to a very massive star in so far as there is a nearly equilibrium 
balance between radiative and gravitational forces.

The effects of heating from the AGN are, for quite different reasons, also 
likely to be independent of absorption (so long as it is not excessive, i.e. 
not Compton thick). \citet{saz05} present a simple analytical exploration 
of the effects and Paper I presents a more detailed one-dimensional
treatment. The photons which dominate the heating process are in the 
moderately hard region ($\sim$50keV), and we know from 
X-ray absorption studies that AGN are typically optically thin 
to such radiation. Metal line resonance absorptions dominate the 
absorption unless the spectrum is extremely hard, and in those cases 
Compton absorption would be dominant. If we consider 
the issue on a per atom basis, all that matters is the heating per atom, 
which scales as $r^{-2}$ (assuming that the fluid is optically thin to hard
X-rays), and the cooling rate per atom which scales as the 
density. Since the latter can also scale as $r^{-2}$ or even falls off at a 
steeper rate in some circumstances, the heating can balance or exceed 
cooling over an extended range of radii. If that happens, the gas 
temperature will rise towards the radiation temperature, $T_X \sim 2 \times
10^7$ K. Then, since this exceeds the virial temperature in almost all 
galaxies, the heated gas, having thermal energy higher than its 
gravitational energy, will be accelerated outwards and tend to drive a 
wind into the surrounding fluid. Of course, since this will shut off the 
accretion flow and the fuel to the central source, the result will be a burst 
of energy output followed by much slower cooling of the shocked gas 
and a repeated burst at a much later time. Thus, episodic accretion is 
expected.

The mechanical energy input is more localized to the vicinity of the 
SMBH and would be efficient in ``protecting" the SMBH from very high 
rates of accretion, except for one important caveat. It necessarily 
happens that such episodes of high rates of energy deposition will 
occur when the central gas densities are high, and under such 
circumstances the gas will tend to radiate away the input energy unless 
forbidden to do so as has occurred in some calculations \citep{boo09}. 
This, as we shall see, makes the energy input rather inefficient
in driving outflows and in protecting the SMBH from excessive
accreation. But the momentum input cannot be radiated 
away, and, as we shall see in the remainder of the paper, it is very 
efficient in limiting the infall and accretion onto the central SMBH.
Mechanical input, via either thermal or momentum based
mechanisms, will also tend to produces episodic accretion.

The broad-line gas outflow must drive a strong shock into the ambient 
gas, and that, in turn, given standard physics, should accelerate 
charged particles efficiently via a variant of the first order Fermi 
process \citep[cf.][]{bla78,bel78,bla87}. Then this relativistic fluid will 
further drive the outflow and, since thermal radiation is suppressed for 
this component, the conversion may somewhat enhance the effects of 
feedback. But, overall, this process simply transforms internal energy 
from one form to another and so, whereas it may be observationally 
quite significant, it will have a relatively small global effect. Two recent 
papers that have explored these processes are \citet{fuj07} and 
\citet{jia10}; see also \citet{sir10}.

\section{The Models}\label{model}
In this section, we summarize the main properties of the galactic models 
adopted in this study. A detailed description of the galaxy models and
input physics is given in CO07, Papers I and III.

We study galaxy models characterized
by the effective radius of the galaxy stellar component $\Re=6.91$ kpc,
an initial stellar mass $M_{*}=2.87\times10^{11} \Msun$, and central
aperture velocity dispersion $\sigma_{\rm a}=260$ $\rm km~s^{-1}$.
This represents approximately the typical $L_{*}$ galaxy which \citet{yu02}
find contain the bulk of the identified mass in SMBHs.
The stellar mass distribution which is embedded in a dark matter halo
is described by the \citet{jaf83} model while the total mass density 
distribution follows a $r^{-2}$ profile; all the relevant dynamical 
quantities need in the simulation are given in \citet{cio09b}. The initial 
SMBH mass is $\Mbh = 2.87 \times 10^8 \Msun$, following the 
\citet{mag98} relations ($\Mbh\sim 10^{-3} M_*$). The simulations are 
for an isolated, giant elliptical galaxy without the effect of the intracluster 
medium, as outflow boundary conditions are set at the galaxy outskirts 
($\sim 250$ kpc), so that the interstellar medium (ISM) is provided by 
the recycled gas produced by stellar evolution. The simulations
starts at 2 Gyr, which corresponds to a redshift of $z\sim3.2$ for the
$\Lambda$CDM cosmology with $\Omega_m =0.3$, 
$\Omega_{\Lambda} = 1-\Omega_m = 0.7$ and $H_0 = 70$ km
s$^{-1}$ Mpc$^{-1}$ and ends at 14 Gyr.

The input physics of the model is fully described in Paper I. Here we 
recall the most important aspects. The instantaneous bolometric 
accretion luminosity is
\beq
\lbh = \epsem \Mdotacc c^2,
\eeq
and we adopt an advection-dominated accretion flow (ADAF)-like
radiative efficiency as
\beq
\epsem = \epszero \frac{A \mdot}{1+A \mdot}, ~~~~\mdot \equiv \frac{\Mdotacc}{\Mdotedd},
\eeq 
where $\Mdotedd = \ledd / \epszero c^2$, and $A$ is a free parameter 
so that $\epsem \sim \epszero A \mdot$ for $\mdot \ll A^{-1}$. We fix 
$A=100$ in our simulations \citep{nar94}, and we adopt for the peak EM 
efficiency $\epszero=0.1$ or $0.2$ consistent with estimates based on 
the \citet{sol82} argument. In the treatment of radiation feedback, we 
consider the radiation pressure as well as heating/cooling feedback, 
including photoionization, Compton and line heating \citep{saz04,saz05}.
In accordance with both observations and theoretical expectation, the
transformation of accreted mass to electromagnetic energy output
declines dramatically at low accretion rates.

In the mechanical feedback treatment, the fiducial instantaneous
mechanical luminosity of the disk wind is calculated as
\beq
\ldw = \epsw \Mdotacc c^2 + \epsii c^2 (1-f_{\rm rem,h}) \frac{M_{\rm dh*}}{\tau_{\rm *h}}
\eeq
where $\epsw$ is the mechanical efficiency of the wind, and the second 
term represents the energy associated with the Type II supernova (SNII) 
explosions of the high-mass stars in the circumnuclear disk (see Paper I, 
Equation (20) for details). Here, $M_{\rm dh*}$ is the current mass in the disc
in high mass ($M>8\Msun$) and $\tau_{\rm *h}$ is their typical lifetime. 
In this work, we restrict attention to the 
commonly assumed case of a constant value of $\epsw$ (e.g. SDMH05), 
which corresponds to Type A models in Papers I and III. Physically, a 
fixed mechanical efficiency implies that the mass accreted by the central 
SMBH has a fixed relation to the mechanical energy flowing out of the 
central regions. We here neglect the jet effects, which are expected to be 
effective only in the low-luminosity, hot accretion phases at late-time 
evolution. The reference models (A0 and A1) from Paper III study the 
evolution of gas and the mechanical feedback from SMBHs, and solve 
Eulerian equations of hydrodynamics with mass, energy, and momentum 
sources (see Paper I).  In order to study the effect of each physical process, 
i.e., mass, energy, and momentum feedback, we build several models 
which neglect one or two of physical terms. We discuss the details of each 
model and comparison of them in the following section.

\begin{table*}
   \begin{center}
   \caption{Summary of One-dimensional Model Properties\label{tab1}}
   \vskip+0.5truecm	
\resizebox{\textwidth} {!}{
  \begin{tabular}{c|c|c|c|c|c|c|c|c||c|c|c|c}\hline\hline
  &  & & & {\footnotesize Bottom}& \multicolumn{4}{|c||}{Feedback} & & & \\ \cline{6-9}
& Model &	$\epsw$ &{\small $\eta (\vwten^{-2})$$^{a}$ }& {\footnotesize Layer}$^{b}$ &{\footnotesize Radiation} & {\footnotesize Energy} & {\footnotesize  Momentum} & {\footnotesize Mass} & log $\Delta\Mbh$ & log $L_X$ & log $l_{\rm BH}^{\rm eff}$$^{c}$ & log $\Delta E_{\rm w}$$^{d}$\\
{\footnotesize(1)} & {\footnotesize(2)} &{\footnotesize(3)}&{\footnotesize(4)}&{\footnotesize(5)}&{\footnotesize(6)}&{\footnotesize(7)}&{\footnotesize(8)}&{\footnotesize(9)}&{\footnotesize(10)}&{\footnotesize(11)}&{\footnotesize(12)} \\
  \hline
1 & A0 $^{e}$ & 	$5 \times 10^{-3}$ & 9 & $\times$ & $\surd$  & $\surd$ & $\surd$ & $\surd$  & 6.72 & 37.11 &  -7.98 &58.67\cr
2 & MA0 $^{e}$ & 	$5 \times 10^{-3}$ & 9 & $\times$ & $\times$& $\surd$ & $\surd$ & $\surd$  & 6.72 & 37.11&  -7.98&58.67\cr
{\bf 3} & {\bf EPM0-R} & 	${\bf5 \times 10^{-3}}$ &{\bf 9} &  ${\bf \surd}$ & ${\bf \surd}$  & ${\bf \surd}$ & {\bf$\surd$} & $\surd$  & {\bf7.07}&{\bf 39.75} & {\bf-10.58} &{\bf59.02}\cr
4 & EPM0 & 		$5 \times 10^{-3}$ & 9 &  $\surd$ & $\times$& $\surd$ & $\surd$ & $\surd$  & 7.13& 37.84 &  -10.58 &59.08\cr
5 & PM0 & 		$5 \times 10^{-3}$ & 9 &  $\surd$ & $\times$& $\times$& $\surd$ & $\surd$  & 7.13&37.84 &  -10.58 &59.08\cr
{\bf 6} & {\bf E0} $^{f}$ & 		${\bf5 \times 10^{-3}}$ & {\bf9 }&  $\surd$ & $\times$& $\surd$ &$\times$& $\times$ &{\bf 10.32}& {\bf41.31} & {\bf-3.61} &{\bf 62.27}\cr
\hline
7 & A1 $^{e}$ & 	$2.5 \times 10^{-4}$ & 0.45 &  $\times$ &$\surd$& $\surd$ & $\surd$ & $\surd$  & 7.38& 36.36 &  -6.72&58.02\cr
8 & MA1 $^{e}$ & 		$2.5 \times 10^{-4}$ & 0.45 &  $\times$&$\times$& $\surd$ & $\surd$ & $\surd$  & 7.52& 38.09 &  -6.48 &58.17\cr
{\bf 9} & {\bf EPM1-R} &  		${\bf2.5 \times 10^{-4}}$ & {\bf0.45} &  $\surd$ &$\surd$& $\surd$ & $\surd$ & $\surd$  & {\bf8.04}& {\bf39.59} & {\bf -8.23}&{\bf 58.69}\cr
10 & EPM1 & 		$2.5 \times 10^{-4}$ & 0.45 &  $\surd$ &$\times$& $\surd$ & $\surd$ & $\surd$  & 7.76&38.82 &  -8.15&58.41\cr
11 & PM1 & 		$2.5 \times 10^{-4}$ & 0.45 &  $\surd$ &$\times$&$\times$& $\surd$ & $\surd$  & 7.76&38.82 &  -8.15 &58.41\cr
{\bf 12} & {\bf E1} $^{f}$ & 		${\bf2.5 \times 10^{-4}}$ & {\bf0.45} &  $\surd$ &$\times$& $\surd$ & $\times$ & $\times$  & {\bf10.33}&{\bf41.29} & {\bf -3.62}&{\bf 60.98}\cr
\hline
13 & EPM2-R & 		$1 \times 10^{-3}$ & 1.8 &  $\surd$ &$\surd$& $\surd$ & $\surd$ & $\surd$  & 7.59& 40.13 &  -9.25 &58.85\cr
14 & EPM3-R & 		$1 \times 10^{-4}$ & 0.18 &  $\surd$ & $\surd$& $\surd$ & $\surd$ & $\surd$  & 8.24& 40.39 &  -7.79 &58.50\cr
15 & EPM4-R & 		$5 \times 10^{-5}$ & 0.09 &  $\surd$ & $\surd$& $\surd$ & $\surd$ & $\surd$  & 8.29& 40.03 &  -7.51 &58.25\cr
16 & EPM5-R & 		$2.5 \times 10^{-5}$ & 0.045 &  $\surd$ & $\surd$& $\surd$ & $\surd$ & $\surd$  & 8.78& 40.33 &  -5.77 &58.43\cr
17 & EPM6-R & 		$1 \times 10^{-5}$ & 0.018 &  $\surd$ & $\surd$& $\surd$ & $\surd$ & $\surd$  & 9.11& 40.50 &  -5.30 &58.37\cr
18 & EPM7-R & 		$5 \times 10^{-6}$ & 0.009 &  $\surd$ & $\surd$& $\surd$ & $\surd$ & $\surd$  & 9.29& 39.97 &  -5.14 &58.24\cr
  \hline\hline
   \end{tabular}}
   \end{center}
{\bf Note.} Model names (except for Model 1, 2, 7 and 8) indicate the 
activated physics (symbol $\surd$) in the simulations as detailed in 
Column 6-9. For example, in E0 and E1 models only mechanical 
energy feedback is allowed, while in PM0 and PM1 models 
momentum and mass are considered, but not mechanical energy.
We adopt $0.2$ for the peak EM efficiency $\epszero$.
\begin{flushleft}
$^{a}$$\eta = 2 \epsw c^2 / \vw^2$ in $\vwten^{-2}$ unit where 
            $\vw=\rm 10,000 km/s$\\
$^{b}$Models with mass, energy and momentum added to the 
            bottom layer.\\
$^{c}$$l_{\rm BH}^{\rm eff} \equiv L_{\rm BH,opt}^{\rm eff} / L_{\rm Edd}$ 
           where $L_{\rm BH}$ is the SMBH luminosity in the optical band
           after absorption, i.e., as it will be seen from infinity.\\
$^{d}$$ \Delta E_{\rm w} \equiv \epsw \Delta \Mbh c^2$, in erg, where 
            $\Delta \Mbh$ is in $\Msun$ units.\\
$^{e}$ These models correspond to the models A0, MA0, A1 and MA1 
            in Papers I and III, but calculated with some difference in the 
            numerical grid spacing.\\
$^{f}$ The model similar to one adopted in SDMH05.\\
\end{flushleft}
\end{table*}

\section{Exploring one-dimensional models}\label{Result}
The model properties and results are given in Table 1. The mechanical
efficiencies $\epsw$ are given in Column 3 and the corresponding 
values of $\eta \equiv 2 \epsw c^2 / \vw^2$ are given in Column 4.  We devote 
Column 5-9 to present ($z=0$) model properties. First, for some models 
indicated with the symbol $\surd$ in Column 5, we distribute the 
mechanical feedback energy (momentum and mass) only at the lower 
boundary of the grids to mimic the common treatment of mechanical 
feedback \citep[e.g. SDMH05,][JNB09]{dim05}. Instead, models indicated 
with the symbol $\times$ in Column 5 have a distributed feedback as in 
Papers I and III where we attempt to estimate the gradual deposition of 
mass, energy and momentum taken from the outflowing wind and going 
into the ambient gas as a function of radius. We then build several models 
which neglect one or 
two of physical process, i.e., mass, energy, and momentum feedback in 
order to study their effects showing the inclusion of each term in Column 
6-9. For example, Model 3 (EPM0-R, in bold face) distributes the 
mechanical feedback only into the bottom layer, and includes the radiation 
feedback and all physical terms, i.e., mass, energy, and momentum, in the 
mechanical feedback. On the other hand, Model 6 (E0) adopts a treatment 
similar to that in SDMH05 as it assumes the same mechanical feedback 
efficiency, only includes the mechanical energy feedback and distributes it 
into the bottom layer of the grid, neglecting the mass and momentum
added back into the flow.

Models 1-6 adopt the standard (high) efficiency  $\epsw=5\times 10^{-3}$, 
as SDMH05 and JNB09 and Models 7-12 assume a factor of 20
lower efficiency, perhaps in better accord with observationally based 
estimates \citep{moe09,ara10} and Models 13-18 adopt other efficiencies to 
show how final properties depend on the assumed mechanical efficiency. 

\subsection{High efficiency models}
\begin{figure}
\centering
\epsfig{file=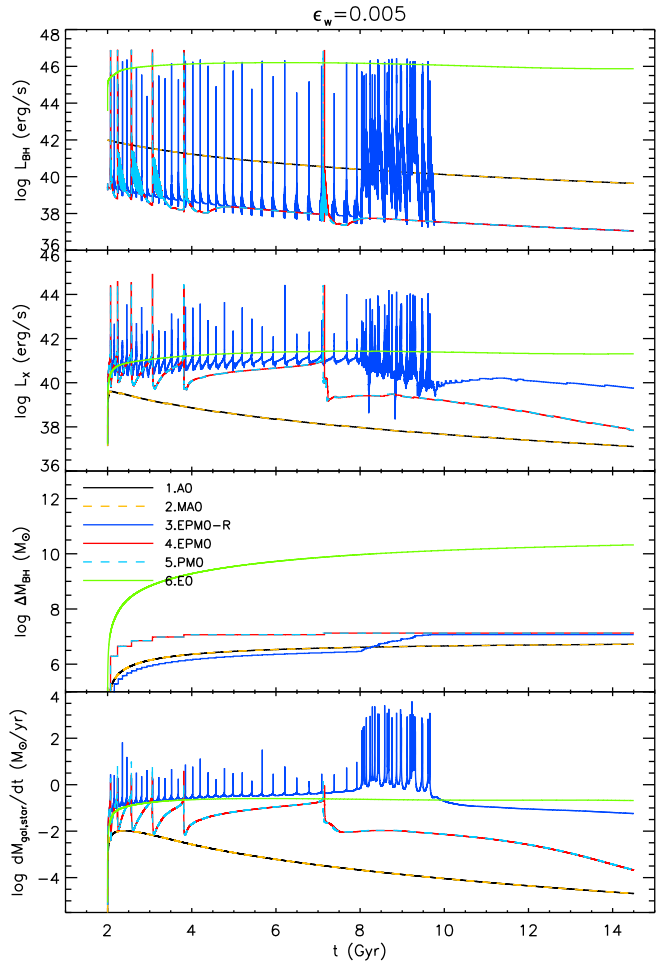,width=\columnwidth}
\caption{Models 1-6 with constant and high mechanical efficiency 
$\epsw=5 \times 10^{-3}$ ($\eta = 9$). From top to bottom, the 
SMBH luminosity, X-ray luminosity, mass accreted on the central SMBH, 
and star formation rate are shown with different line types and colors as 
indicated in the third panel. Note how the model that excludes 
momentum feedback, ``E0", has by far the highest growth of the central 
SMBH.
\label{fig:A0}}
\end{figure}

To mimic the common treatment (e.g., SDMH05, JNB09), we build the 
Model 6 (E0) that only includes the Mechanical energy feedback
with the standard (high) efficiency $\epsw=5\times 10^{-3}$. 
In this model, we estimate the mass inflowing to the SMBH, convert
it to energy with the given efficiency, and add this energy only into the 
bottom layers of the surrounding gases. For comparison, Model EPM0-R 
has identical efficiency but adds also mass and momentum to the bottom 
layers using Equations (5a-d) with $\eta=9$, as appropriate for the chosen 
efficiency and wind velocity of $\vwten=1$. These two models are shown 
as blue and green lines in Figure~\ref{fig:A0}. We see that allowing for 
momentum and mass feedback reduces the black hole growth by a factor 
of 1000. The more consistent model has a much lower final X-ray 
luminosity and final SMBH Eddington ratio. The effect of including or not 
including radiative heating is relatively minor, as can be seen by 
comparing Models 3 and 4 or 1 and 2. Also the mechanical energy 
feedback is considerably less important (as expected) than the 
momentum input, as can be seen by comparing Models 4, 5, and 6.
Finally it might be thought that some of the effects observed in these
comparisons are due to the change from Paper I of adding the feedback
to the bottom layers alone in the present simulations, rather than over a 
distributed range of radii to mimic the effects of due to a broad line wind. 
But comparison between Models 1 and 3, where there are only small 
differences (and Model 1 is identical to A0 of Paper III), shows that the 
differences which may be attributed to distributed feedback are small.
In summary, examination of Models 1-6 shows that including 
momentum drastically increases the effects of feedback.

\subsection{Low efficiency models}
\begin{figure}
\centering
\epsfig{file=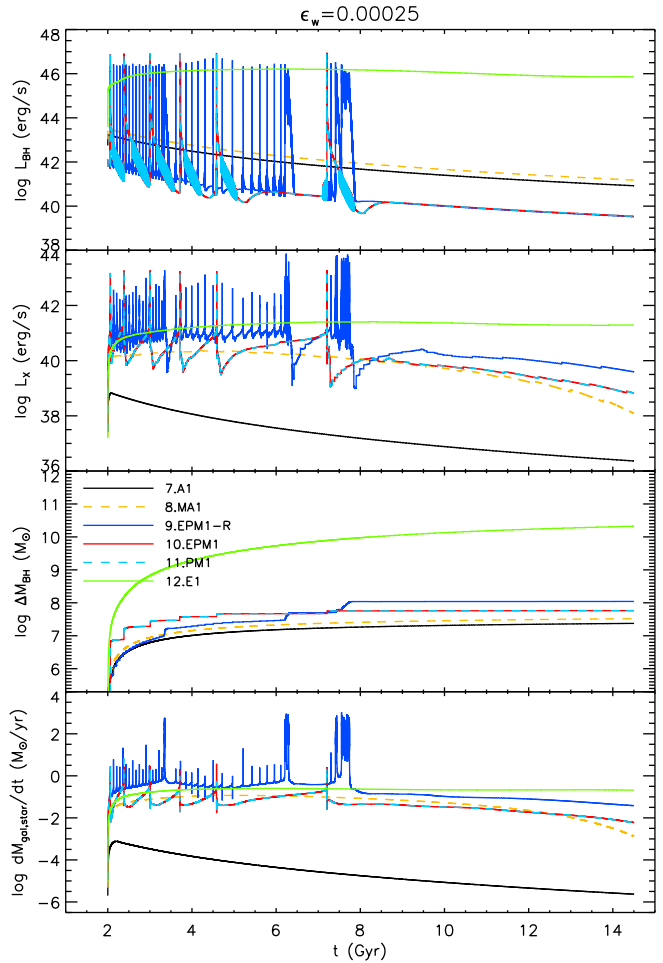,width=\columnwidth}
\caption{Models with constant and low (observation based) mechanical 
efficiency, $\epsw=2.5 \times 10^{-4}$ ($\eta = 0.45$). From top to bottom, 
the SMBH luminosity, X-ray luminosity, mass accreted on the central 
SMBH, and star formation rate are shown, colors and line types as in 
Figure~\ref{fig:A0}. Again momentum feedback is the most important 
physical process in protecting the central SMBH from excessive growth.
\label{fig:A1}}
\end{figure}

\begin{figure}
\centering
\epsfig{file=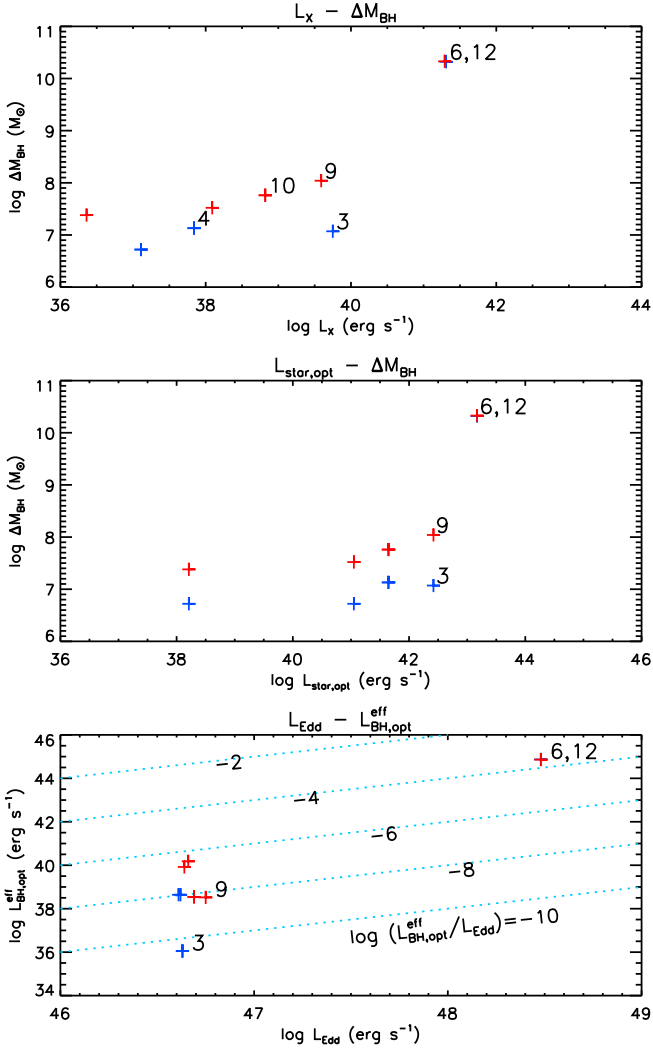,width=\columnwidth}
\caption{Distribution of all models in the mass-luminosity diagram 
measured at z=0, where different colors show different wind efficiencies 
and $\eta$ values ($\epsw=5 \times 10^{-3}$ and $\eta =9$ for blue,
$\epsw=2.5 \times 10^{-4}$ and $\eta =0.45$ for red). The distribution of
models in the Eddington luminosity - effective SMBH luminosity plane is 
shown in the bottom panel. Five diagonal lines (from top to bottom) show 
$l_{\rm BH}^{\rm eff}=L_{\rm BH, opt}^{\rm eff} / L_{\rm Edd}$$=10^{-2}$, 
$10^{-4}$, $10^{-6}$, $10^{-8}$, $10^{-10}$ respectively.
\label{fig:m_l}}
\end{figure}

Next, we turn our attention to Models 7-12 which have a much lower
mechanical efficiency than typically assumed and it is at a level better in 
accord with existing (and highly imperfect) observational indications. The 
value for the dimensionless parameter $\eta$ in these cases is only 0.45 
(i.e. of order unity), so that we expect that inclusion or exclusion of the 
mass and momentum input will make relatively less difference. What do 
we find? In fact, the differences are reduced by about half an order of
magnitude (0.5 dex), but it remains true that the Model 12 (like Model 6), 
without either momentum feedback or radiation, has an unacceptably 
large growth of the central SMBH and an unacceptably large final SMBH 
luminosity, as shown in Figure~\ref{fig:A1}. Models 6 and 12 also show 
thermal X-ray emission greater than $10^{41}$ erg/s, which is on the 
upper side of what is typically observed in normal elliptical galaxies.

We summarize the properties of Models 1-12 in Figure~\ref{fig:m_l} 
showing the present-day (14 Gyr) SMBH mass in solar mass versus 
X-ray gas and optical stellar luminosities. We show the high efficiency 
models (Models 1-6) in blue and the low efficiency models (Models 6-12) 
in red. The fiducial models, Models EPM0-R and EPM1-R with mass and 
momentum feedback, and Models E0 and E1 that only include energy 
feedback, are marked with their model numbers. As discussed above, 
including the momentum and mass feedback not only significantly 
reduces the SMBH growth but also results in a much lower final X-ray 
luminosity and final SMBH Eddington ratio. 

\subsection{Wind efficiency dependence of bottom-layer models}
\begin{figure}
\centering
\epsfig{file=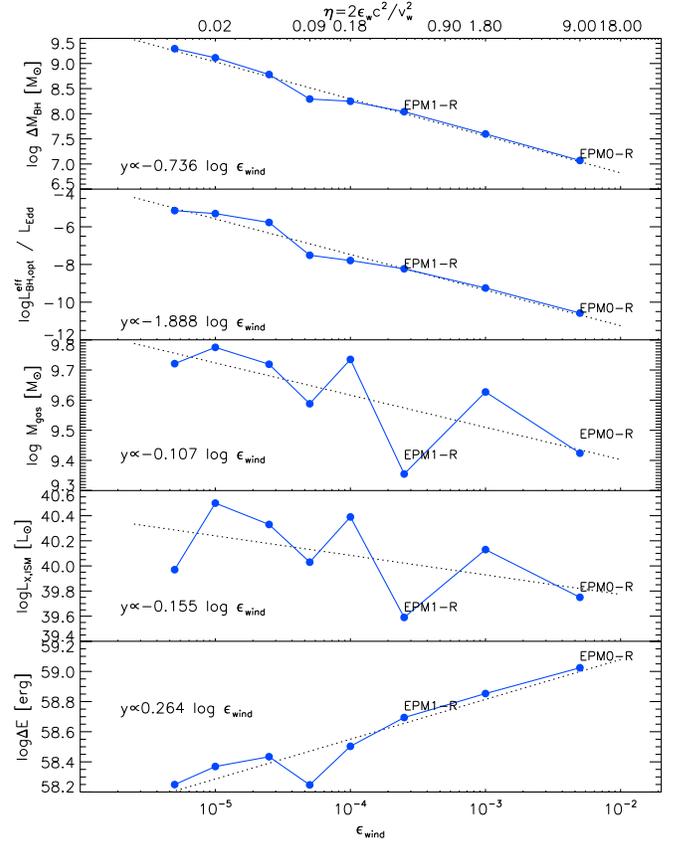,width=\columnwidth}
\caption{Dependencies of present-day, global quantities of EPM\#-R 
models in Table 1, as a function of mechanical efficiency $\epsw$. From 
top to bottom, the SMBH mass growth, BH Eddington ratio, galaxy gas 
mass inside 10 $\Re$, X-raygas luminosity, and total wind feedback 
energy are shown. The linear fits to the data are shown in dotted lines, 
and the fitting results are shown in each panel. As expected, the 
assumption of a higher wind energy efficiency does correspond to 
greater feedback effects, but at a much less than linear rate.
\label{fig:epsw}}
\end{figure}

We test eight different values of $\epsw$ for models ranging from 
$5\times10^{-6}$ to $5 \times 10^{-3}$ for models with the bottom-layer 
treatment and all feedback physics activated (i.e. Models EPM\#-R). 
These models correspond to Models 3, 9, and 13-18 in Table 1. We 
summarize the results at the epoch of 14 Gyr in Figure~\ref{fig:epsw},
where the least square linear fits of several global quantities of interest 
are also given. Note here that the growth of SMBH mass and the BH 
luminosity Eddington ratio are decreasing functions of $\epsw$. If the 
feedback efficiency is low, too much mass is accreted to the central 
SMBH, as expected. In the case of the gas mass and the predicted 
X-ray luminosity of the hot ISM, they are decreasing functions with 
increasing wind efficiency but with large scatters. We also show the total
 mechanical feedback energy (i.e. $ \Delta E_{\rm w} = \epsw \Delta \Mbh 
 c^2$) in the last panel, which increases as the wind efficiency increases.
In this case as in several of the others, while the behavior is 
approximately monotonic in the direction expected, the approximate 
power law index is much less than unity, since larger efficiency gives a 
larger value for our dimensionless parameter, $\eta$, and thus a smaller 
fraction of the inflowing gas is actually accreted onto the central SMBH.

\subsection{The effect of mass removal from the circumnuclear disk on 
                      purely radiative models}
\begin{figure}
  \centering
  \includegraphics[width=\columnwidth]{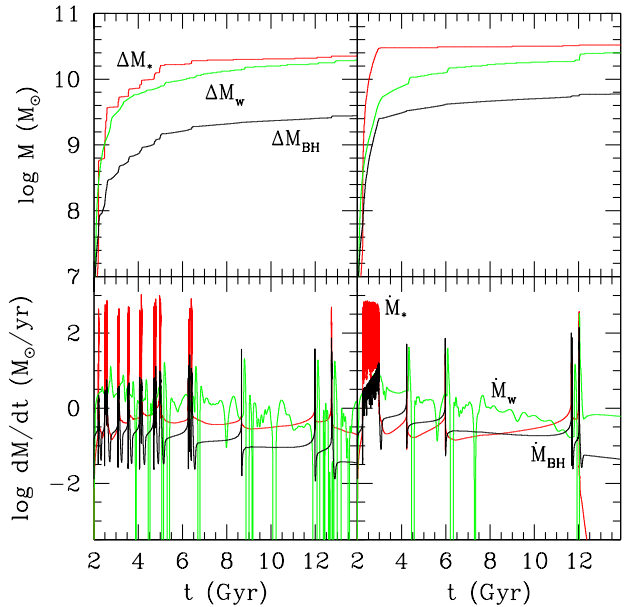}
  \caption{Time evolution of relevant mass budgets (top panels) and 
  corresponding mass rates (bottom panels) in two purely radiative 
  models (without mechanical feedback), differeing in the treatment 
  of the circumnuclear disk mass budget. The model on the left 
  panels is Model RB0 (see Table 2 in Paper I), while the model in the 
  right panels is identical in all its properties to RB0, except that no mass
   is lost by the circumnuclear disk. Top panels: total mass accreted by 
   the central SMBH (black), of the total mass of ISM ejected at 
   10 $R_{\rm e}$ (green), and of the total mass in new stars 
   accumulated within 10 $R_{\rm e}$ (red). Bottom panels: the 
   corresponding mass rates are identified by same colors as in top 
   panels. The gas production of the passively evolving stellar population 
   steadely declines from $\approx 10 M_{\sun}$/yr at the beginning 
   down to less than $1 M_{\sun}$ at the end.
  \label{fig:rb0nm}}
\end{figure}

In line with the present exploratory discussion, it is of some interest
also to check the effects of different amounts of mass removal from the
circumnuclear disk via disk wind, in the case of purely radiative models. 
In fact, we recall that in the purely radiative models presented in Paper I 
(such as model RB0 in Table 2 therein) we do not add mechanical 
feedback to the equations of hydrodynamics, but the mass, momentum 
and energy fluxes of the nuclear wind (and of the jet) are nonetheless 
computed, in order to satisfy Equations \ref{eq:Mdot}--\ref{eq:sol} for 
assigned mechanical efficiency and fiducial nuclear wind velocity.

Therefore, purely radiative models depend indirectly on the assumed 
mechanical efficiency, with high-efficiency models ejecting a larger 
fraction of the gas from the circumnuclear wind, and therefore reducing 
the amount of gas available for accretion on the SMBH. Here we compare 
the evolution of the purely radiative model RB0 in Paper I (a model with 
radiative efficiency $0.1$ and with high constant mechanical efficiency 
$5 \times 10^{-3}$), with an identical purely radiative model, in which the 
mechanical efficiency has been reduced to zero, therefore excluding 
mass loss from the circumnuclear disk.

The situation is illustrated in Figure~\ref{fig:rb0nm}, where the left panels 
refer to model RB0, and the right panels to the model without mass ejection 
from the nuclear disk. In the top panels we show the time evolution of the 
total mass accreted by the central SMBH (black line), of the total ISM mass 
ejected by the galaxy as a galactic wind (green line), and finally of the 
accumulated mass in new stars (red line). In the bottom line, the 
corresponding rates are shown and identified with the same colors.

Unsurprisingly, the SMBH grows significantly more (by
a factor of $\sim 2$) in the model RB0 without nuclear wind mass loss
($\log \Delta M_{\rm BH}/M_{\odot} \simeq 9.78$) than in the model
 with mass ejection ($\log \Delta M_{\rm BH}/M_{\odot}\simeq 9.45$).
The major difference in the accretion history of the two models is 
particularly evident in the first Gyr of evolution, when large amounts 
of gas flow on the central region of the galaxy. Note how the SMBH mass 
of the model without nuclear mass ejection (right panels) reaches a 
value similar to the SMBH mass of model RB0 (left panels) at the end 
of the simulation. As a consequence, the gas near the SMBH is 
gravitationally more bound in the first model - especially at early times 
when the mass losses are significant. As can be seen, the star 
formation history in the two models is almost parallel to their SMBH 
accretion, and the larger radiative energy output in the model without 
nuclear mass ejection is accompanied by a larger starburst at early 
times, with a final mass of new stars of $\log \Delta M_{*}/M_{\odot} 
\simeq 10.5$ (red lined), to be compared with $\log\Delta M_{*}/M_{\odot} 
\simeq 10.36$ in RB0 model without disk mass ejection. Finally, 
consistently with the larger energy input of the model shown in the 
right panels, the galactic wind expelled a total ISM mass of $\log \Delta 
M_{\rm w}/M_{\odot}\simeq 10.4$ in the model without the disk wind, to 
be compared to $\log \Delta M_{\rm w}/M_{\odot}\simeq 10.3$ in RB0 
model. 

Again, this very simple experiment shows how different treatments in 
the mass balance equations used to describe SMBH can leads to 
significantly different evolutionary histories \citep[cf. also][]{sok05}.

\section{Two-dimensional Model Comparison}\label{Result2d}
\begin{figure}
  \centering
  \includegraphics[width=\columnwidth]{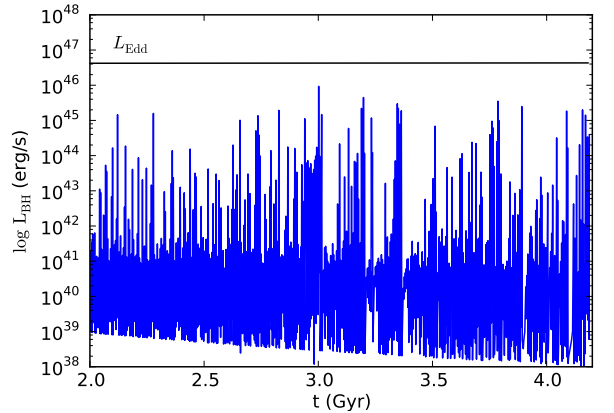}
  \includegraphics[width=\columnwidth]{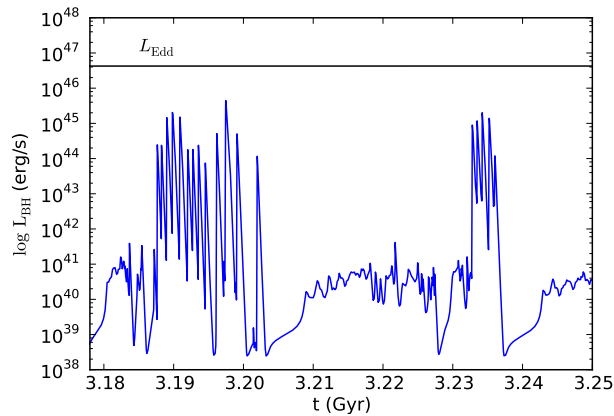}
  \caption{Luminosity versus time for an axisymmetric A model with
    $\epsilon_M = 2.5\times 10^{-4}$.  Above, the AGN luminosity for
    half a Gyr.  Below, the AGN luminosity plotted for a shorter time
    showing the highly variable nature of the accretion events in two
    dimensions.  The BH accretion is much more stochastic than the
    one-dimensional case, but the distribution of Eddington ratios is
    quite similar.}
  \label{fig-twod-l-t}
\end{figure}

\begin{figure*}
 \centering
 \includegraphics[width=\textwidth]{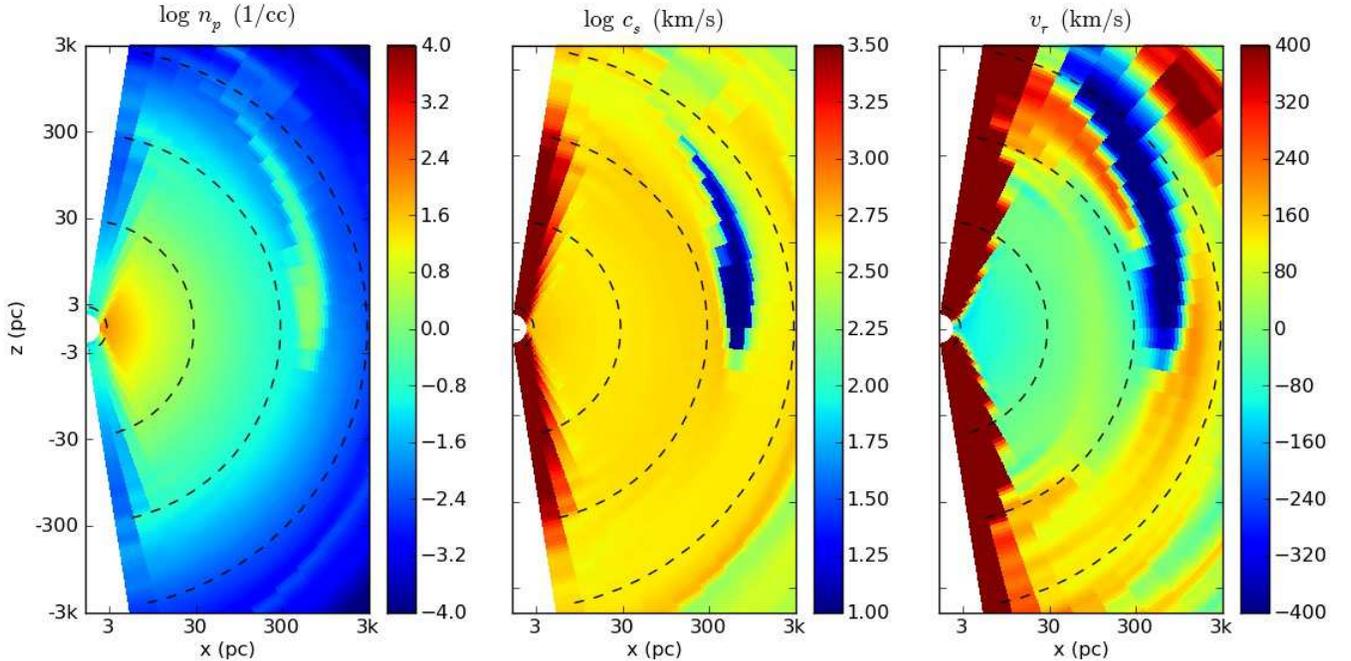}
 \caption{A snapshot from an axisymmetric simulation showing a cold
   blob falling to the center of the galaxy.  On the left, log gas
   density in number of protons per cubic centimeter.  In the center,
   log sound speed in kilometers per second.  On the right, the
   radial velocity in kilometers per second.  The $x$ and $y$ axes
   are logarithmic in the distance to the SMBH.  The cold gas
   was produced by enhanced cooling in an overdense quasi-spherical
   shell with a covering fraction of about one third of the sphere.
   The gas quickly collapses to a ring with a small covering fraction
   and/or fragments as it freely falls to the center of the
   simulation.  }
 \label{fig-twod-rt}
\end{figure*}

\begin{figure}
  \includegraphics[width=\columnwidth]{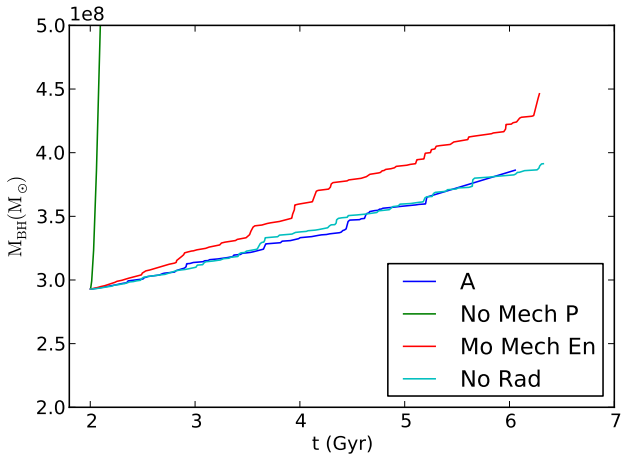}
   \caption{Black hole mass as a function of time for A models with
     $\epsw = 10^{-3}$.  The blue line is the fiducial case with
     all physics included.  The green line has mechanical momentum
     injection turned off so that mechanical feedback is purely via
     energy; the SMBH grows prodigiously, indicating that mechanical
     momentum feedback is by far the dominant process in limiting the
     growth of the black hole.  The red line has mechanical energy
     injection turned off, leaving mass and momentum injection
     unchanged; the SMBH grows somewhat more than the fiducial case,
     indicating that energy feedback plays some role in limiting SMBH
     growth, albeit a sub-dominant one.  The cyan line has
     $\epsilon_{\rm EM}$ set to zero so that there is no energy or
     momentum feedback due to radiation from the central SMBH; this is
     indistinguishable from the fiducial case, indicating that
     radiative feedback plays essentially no role in limiting SMBH
     growth.}
  \label{fig-twod-mbh-t}
\end{figure}

One dimensional models continue to be very useful in establishing the
basic physical processes that are relevant for AGN feedback in giant
elliptical galaxies.  However, one dimensional models are not able to
capture important properties of the actual systems, including the
convective, Rayleigh-Taylor, and Kelvin-Helmholtz instabilities.  One
dimensional models must also rely on a parameterization of the global
deposition of mass, energy, and momentum via the disk wind, while
higher dimensional models are able to simulate the evolution of the
wind self-consistently. We discuss below two-dimensional models
where we have taken exactly the same galaxy model and feedback
characteristics to allow comparisons that are easy to understand.

There have been many numerical simulations of BH accretion and the
subsequent effects on the galaxies containing resulting AGN.  However,
efforts to date divide into three categories.  \citet{dim05},
\citet{deb09}, and \citet{joh09} are examples where the
simulations cover length scales from $\simeq$ 100 pc to tens of kpc
and timescales from a fraction of a Myr to several Gyr.  Galactic
length and timescales are resolved, but the BH accretion and feedback
processes are considered to be sub-resolution.  \citet{kur09b,kur09c}
are examples of multi-dimensional
simulations that cover the length scales from a few AU to $\simeq$ 1
pc.  Length and timescales relevant to BH accretion are resolved, but
these simulations do not approach approach galactic length or
timescales, and infall rates are taken as given.  Finally,
\citet{hop09} and \citet{lev08} are examples of a
multi-resolution studies of BH accretion involving progressively
higher spatial resolution simulations run for progressively shorted
times.  The highest spatial resolution simulations go down to a
fraction of a pc, run for about one mega year of simulation time.
These simulations spatially resolve the accretion process, but do not
reach galactic timescales.  Therefore they cannot self-consistently
calculate the effect of AGN feedback on the gas in the galaxy as a
whole and subsequent BH accretion.

The present work is the only attempt of which we are aware to
simultaneously resolve the inner length scales relevant to BH
accretion (a few pc), outer length scales relevant to galaxies (tens
of kpc), inner timescales relevant to BH accretion (a few years), and
outer timescales relevant to galaxies and stellar evolution (10 Gyr).
However, the region inside of 1 pc including the disk and BH itself
are still treated as sub-resolution physics.

A full description of the two-dimensional simulations and an analysis
of the similarities and differences between the one- and
two-dimensional models is forthcoming.  Briefly, we use the Zeus
hydrodynamics code \citep{sto92} in spherical coordinates with
log-spaced radial bins with $\Delta r/r = 0.1$.  We have extended the
code to include appropriate mass, energy, and momentum source terms
corresponding to stellar evolution, star formation, type 1a and type
II supernova feedback, radiative and mechanical feedback from AGN
activity.  See CO07, Papers I and III, \citet{saz05} for a full
description of the input physics, which are carried over in all
respects except that we have omitted the radiation pressure on dust.
We require the cells to have an aspect ratio of one, giving 30 angular
cells.  Resolution studies have shown little difference in the SMBH
accretion as a function of time as long as the opening angle of the
disk wind is resolved.

The major differences between the one-dimensional code and the
two-dimensional code are in the way that the two codes handle angular
momentum and the disk wind from the AGN.  

The one-dimensional simulations did not permit the simulated gas to
have nonzero angular momentum.  The 2D simulations assume axisymmetry,
but compute the velocity in the $\phi$ direction.  We must assume an
angular momentum profile.  In the present simulations, we avoid
forming a rotationally supported gas disk by choosing the radius of
centrifugal support to be inside the innermost grid cell.  This allows
us to avoid specifying an ad-hoc prescription for angular momentum
transport.  

The net specific angular momentum of the stars providing gas in the
simulation is assumed to be:
\begin{equation}
  \frac{1}{v_\phi} = \frac{1}{f \sigma} + \frac{R}{j} + \frac{d}{\sigma R}
\end{equation}
where $R$ is the distance to the $z$ axis.  This parameterization
gives solid body rotation at small radii and constant specific angular
momentum at large radii.  The first term prevents the rotational
velocity from exceeding $f \sigma$---at intermediate radii, there may
be a region with constant velocity.  When gas is created in the
simulation by stellar evolution, it is given this angular momentum
profile.  The subsequent evolution of the gas velocity on the
computational grid is governed the standard fluid dynamics
conservation laws.

The one-dimensional code employs a phenomenological model to determine
the radius at which energy, mass and momentum from the AGN-driven disk
wind are deposited in the simulation grid.  This model depends on an
assumed instantaneous jet opening angle.  The two-dimensional code
also requires an assumption about the angular dependence of the
energy, mass, and momentum injected by the disk wind at the edge of
the simulation grid.  Once conserved quantities have entered the
simulation grid, the two-dimensional code self-consistently calculates
the time evolution of the material from the disk wind; a separate
phenomenological model is not required.  

For the $A$ models, the opening angle of the jet is chosen so that the
disk wind covers $\pi$ steradians, giving a linear opening half-angle
of $\cos^{-1} \frac{3}{4} \simeq 41^\circ$.  The opening angle does
not depend on the BH luminosity in the $A$ models.  The 1D models
simply require the jet opening angle as a parameter, but the 2D models
require that the flux of material be fully specified as a function of angle
from the $z$ axis.  We use 
\begin{equation}
  \frac{dq}{d\Omega \, dt} \propto \cos^2(\theta)
\end{equation}
where $q$ is mass, energy, or radial momentum, $\Omega$ is solid
angle, and $\theta$ is the angle from the $z$ axis.  This
parameterization gives half of input material within a half opening
angle of $\simeq 41^\circ$.

For the present purpose, the primary result from the two-dimensional
models is that the qualitative conclusions already drawn from one-dimensional
models remain valid.  The dominant physical mechanism regulating black
hole growth is momentum injected by the broad-line wind.  The energy
provided by the mechanical wind has a noticeable but comparatively
small effect.  The effect of other feedback mechanisms is much
smaller than either the mechanical momentum or mechanical energy.  

Figure \ref{fig-twod-l-t} shows the AGN luminosity versus time for one
of the two-dimensional models with a mechanical efficiency of
$\epsw = 0.001$ (corresponding to the one-dimensional model
EPM2-R).  The primary difference in the SMBH growth between the
one- and two-dimensional models is that the two-dimensional models
have much more stochastic growth.  There quiescent periods are not as
quiescent, and the spacing between the major bursts is not as regular in
time.  Both of these are due to instabilities present in multiple
dimensions: quasi-spherical shells of cold gas are able to fragment
and fall into the center bit by bit rather than as a single large
shell.  In the one-dimensional simulations, bursts of accretion form a
hot central bubble that is able to prevent further accretion until the
hot bubble cools---this often leads to very regular spacing of
accretion events in time.  

In two-dimensional simulations, a similar hot bubble is formed, but
cold gas is able to reach the center via Rayleigh-Taylor and
convective instabilities.  An example of this is shown in
Figure~\ref{fig-twod-rt}.  Hot gas simply moves out of the way leading
to much more stochastic SMBH accretion with bursts much more closely
spaced in time.

Figure \ref{fig-twod-mbh-t} shows the SMBH mass versus time for
several two-dimensional simulations where each physical process is
turned off in turn, allowing us to identify which ones are negligible
and which ones play a dominant role in regulating SMBH growth.
Without mechanical momentum injection, the SMBH grows in a
fashion only limited by $L_{\rm Edd}$.  Without
mechanical energy injection, the SMBH grows about a factor of two
faster than the fiducial case.  Mechanical energy plays a role, but it
is much less important than mechanical momentum input.

Turning off {\em all} radiative feedback processes by setting
$\epsilon_{\rm EM}=0$ has little effect on the SMBH
growth.  Making this choice eliminates gas heating as computed by the
expressions in \citet{saz05}, momentum provided by the absorption
of those same photons, as well as momentum provided by electron
scattering that determines the Eddington limit.  The code does not
impose the Eddington limit---it allows the accretion to be limited
self-consistently by adding the Eddington force to the momentum
equation.  Therefore setting $\epsilon_{\rm EM}=0$ means that the SMBH
would not be limited by radiative momentum.  In spite of this, the
mechanical feedback is able to keep the accretion rate to physically
plausible values. The actual optical depth in our simulation for electron
scattering is typically small compared to unity. This is consistent with
observations which show that only a minority of AGNs are ``Compton
thick".

\section{Discussion}\label{Discuss}
The primary purpose of this paper is to quantitatively show, based on 
one- and two-dimensional computations, exactly which processes are 
most important during AGN feedback episodes; which processes are 
most useful in protecting the central SMBH from excessive mass growth 
and which have most effect on the ambient galaxy.  After a central 
outburst, the mechanical energy must be communicated to the ambient 
gaseous fluid by a wind and we, in fact, see these winds in luminous 
galaxies labeling them ``broad-line regions" with outflow velocities 
observed to be $\sim 10,000$ km/s covering conical regions subtending 
20-25 \% of the sky. These winds {\it must} carry mass and radial 
momentum to the ambient fluid, reducing thereby the mass deposited on 
the central SMBH and adding a driving component which cannot  be 
reradiated away by thermal processes. Equations 5a-d summarize
the physics, with the dimensionless parameter $\eta$, indicating the 
importance of mass and momentum outflows.

In the case of an assumed high mechanical efficiency ($\epsw=0.005$), 
we find that, if we suppress the mass and momentum input, then the 
SMBH grows by over a factor of 100 more than if momentum and mass 
flux were properly included in the calculation, and reaches 
masses~$> 10^{10} \Msun$ in both one- and two-dimensional 
calculations. Turning on or off the energy input has relatively much less 
effect, altering the SMBH growth by roughly a factor of 2. Ignoring the 
mass and momentum feedback inputs also leaves the galaxy with a
central optical luminosity from the AGN which is orders of magnitude 
brighter than is seen in nearby elliptical systems.

Compared to these dramatic effects, the uncertainty due to not knowing 
accurately the wind efficiencies has a relatively minor effect. Reducing
the efficiency by a factor of 20 from $5\times10^{-3}$ to 
$2.5 \times 10^{-4}$ reduces the wind energy output by only a factor of 2 
(to $10^{58.7}$ erg) and reducing the efficiency by another factor of 10 
reduces the wind energy output by only another factor of 2.

We also found that redirecting much of the inflowing mass into a BAL 
wind has, by itself, an important effect on models with only radiative 
feedback. In those computations, which do not allow for the redirection, 
the central SMBH again grows far too much in both the one- and 
two-dimensional computations.

In summary, it eventuates that enforcing the conservation of mass,
momentum and energy provides extremely useful constraints in 
estimating the growth of central SMBHs and the feedback effects on the 
surrounding galaxies.

\acknowledgements
We benefited from useful conversations with Lars Hernquist, Thorsten 
Naab, Eve Ostriker and Elliot Quataert. J.P.O. and E.C. acknowledge the 
support of NSF grant AST-0707505. G.S.N. acknowledges the support of 
the Princeton University Council on Science and Technology.

\end{document}